\documentclass[referee]{aa}
\usepackage{txfonts}
\usepackage{graphicx}

\def\msun{\hbox{M$_\odot$}}

\begin{document}

\title{On the structure of Small Magellanic Cloud star clusters}

\author{Andr\'es E. Piatti\inst{1,2}\thanks{\email{andres.piatti@unc.edu.ar}}}

\institute{Instituto Interdisciplinario de Ciencias B\'asicas (ICB), CONICET-UNCUYO, Padre J. Contreras 1300, M5502JMA, Mendoza, Argentina;
\and Consejo Nacional de Investigaciones Cient\'{\i}ficas y T\'ecnicas (CONICET), Godoy Cruz 2290, C1425FQB,  Buenos Aires, Argentina\\
}

\date{Received / Accepted}

\abstract{It has been recently shown from observational data sets
the variation of structural parameters and internal dynamical evolution of star
clusters in the Milky Way and in the Large Magellanic Cloud (LMC), caused by the different  
gravitational field strengths that they experience. We report here some hints for
such a differential tidal effects in structural parameters of star clusters in the
Small Magellanic Cloud (SMC), which is nearly 10 times less massive than the
LMC. A key contribution to this study is the consideration of the SMC as a
triaxial spheroid, from which we estimate the deprojected distances to the
SMC center of the statistically significant sample of star clusters analyzed.
By adopting a 3D geometry of the SMC, we avoid the spurious effects caused
by considering that a star cluster observed along the line-of-sight is
close to the galaxy center. When inspecting the relationships between the
star cluster sizes (represented by the 90$\%$ light radii), their eccentricities,
masses and ages with the deprojected distances, we find: (i) the star cluster
sizes are not visibly affected by tidal effects, because relatively small and large 
objects are spread through the SMC body. (ii) Star clusters with large
eccentricities ($\ge$ 0.4) are preferentially found located at deprojected distances
smaller than $\sim$ 7-8 kpc, although many star clusters with smaller eccentricities 
are also found occupying a similar volume. (iii) Star clusters more massive than
log($M$ /$M_{odot}$) $\sim$ 4.0 are among the oldest star clusters, generally
placed in the outermost SMC region and with a relative small level of flattening.
These findings contrast with the more elongated, generally younger, less massive
and innermost star clusters.}
 

\titlerunning{SMC star clusters}

\authorrunning{Andr\' es E. Piatti }

\maketitle

\markboth{Andr\' es E. Piatti: }{SMC star clusters}

\section{Introduction}

The structure of star clusters evolves over their lifetime, mainly because
of the stellar evolution, two-body relaxation and tidal effects caused by the host galaxy's
gravitational field \citep[e.g.,][]{hh03,lamersetal2005a,km2009,gielesetal2011,webbetal2013,webbetal2014,shukirgaliyevetal2018}. Although mass loss due to  tidal heating has long been treated theoretically and
from numerical simulations  \citep[e.g.,][]{gnedinetal1999,bm2003,gielesetal2006,lg2006,gielesetal2008,kruijssenetal2011,gr2016}, 
the magnitude of this phenomenon on different star clusters has been
more difficult to measure. Indeed, the observed
variation across the body of a galaxy of  the core, half-mass, and Jacobi radii, cluster eccentricity,  half-mass 
relaxation time, cluster  mass, among other star cluster parameters, has relatively recently
been studied in some limited number of cases.

\citet{piattietal2019b} analyzed the extent in shaping the structural parameters and
internal dynamics of the globular cluster population caused by the effects of the
Milky Way gravitational field. They employed a homogeneous, up-to-date data set 
with kinematics, structural properties, current and initial masses of 156 globular 
clusters, and found that, in overall terms, cluster radii increase as the Milky Way potential weakens. Core radii increase at the lowest rate, while Jacobi radii do at the fastest one,
which implies that the innermost regions of globular clusters are less sensitive to 
changes in the tidal forces with the Galactocentric distance. The Milky Way gravitational
field also differentially accelerates the internal dynamical evolution of globular clusters, 
with those toward the bulge appearing dynamically older. Globular clusters with
large orbital eccentricities and inclination angles experience a higher mass loss 
because of  more tidal shocks at perigalacticon and during disc crossings
\citep{piatti2019}. 

Milky Way open clusters are also  subject to tidal heating. Because they are younger
than globular clusters, mass loss due to stellar evolution can be more important,
particularly if they are younger than few  hundred million years, while two-body relaxation becomes important as the mass loss rate due to stellar evolution continues to decrease 
\citep{lamersetal2005a}. Nevertheless, shocks with giant molecular clouds are known 
to be the dominant source of mass-loss over the open cluster’s lifetime \citep{lg2006}.
\citet{joshietal2016} studied a statistically complete sample of open clusters
located within 1.8 kpc from the Sun and found that their present-day masses follow
a linear relationship with their respective ages. Assuming that the gravitational field
does not vary significantly within such a circle,  stellar evolution could be responsible
for such a trend.

\begin{figure*}
\includegraphics[width=\textwidth]{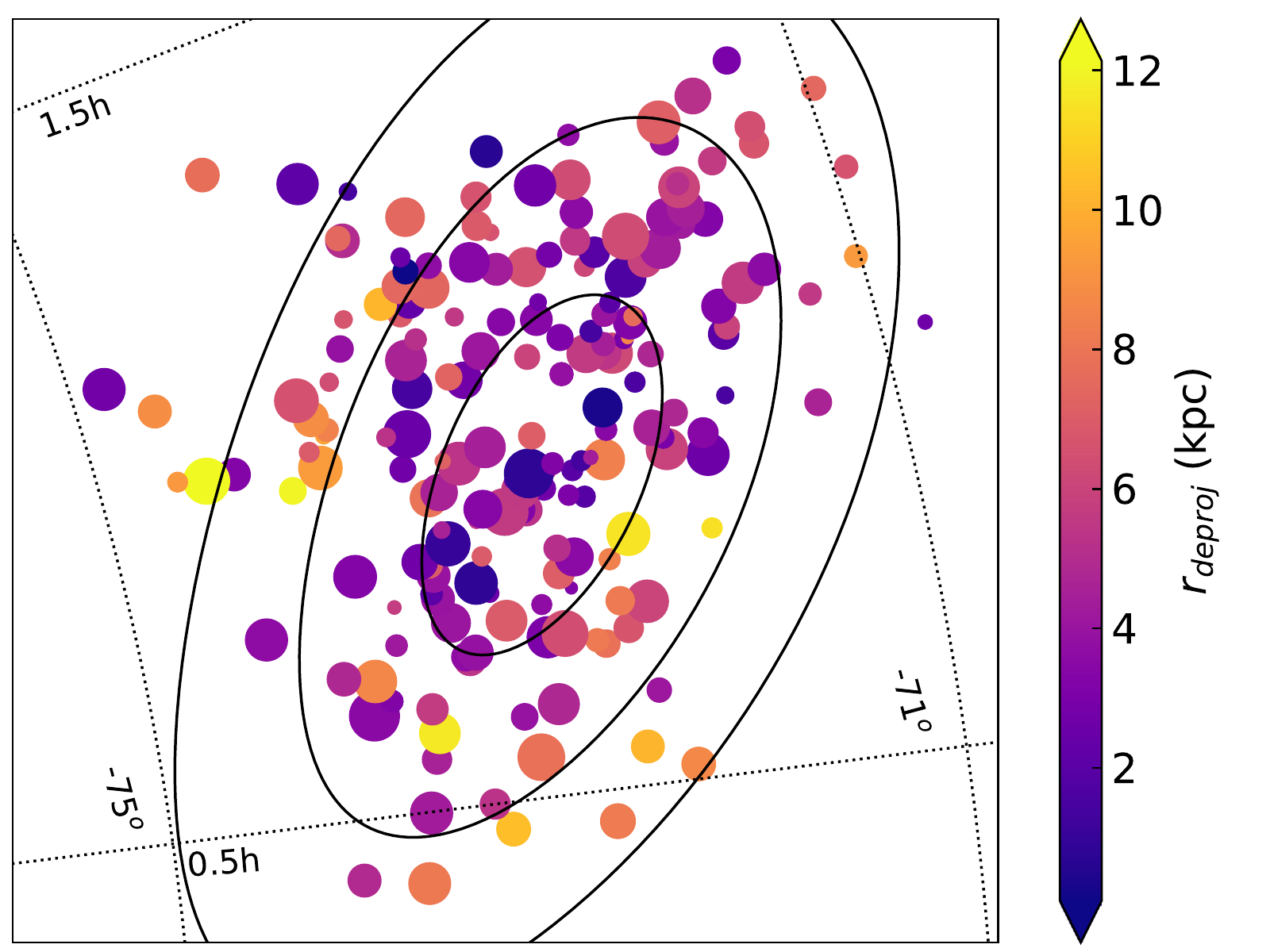}
\caption{Equal-area Hammer projection of the SMC in equatorial coordinates. Three
ellipses with semi-major axes of 1$\degr$, 2$\degr$, and 3$\degr$ are superimposed.
Symbols are colored according to the star cluster distance to the SMC center, while
their sizes are proportional to the star cluster 90$\%$ light radii.}
\label{fig:fig1}
\end{figure*}

The Large Magellanic Cloud (LMC) is nearly 10 times less massive than the
Milky Way \citep{deasonetal2020} and differential tidal effects are also seen
within its population of globular clusters. \citet{pm2018} built extended stellar density and/or surface brightness radial profiles for almost all the known LMC
globular clusters and found that those located closer than $\sim$ 5 kpc from the LMC
center contain an excess of stars in their outermost regions with respect to the stellar density expected from a King profile, which are not seen in globular clusters located
beyond $\sim$ 5 kpc from the LMC center. In addition, globular cluster sizes show
a clear dependence  with their positions in the galaxy, in the sense that the closer
the globular cluster to the LMC center, the smaller its size. Although the masses
of the LMC globular clusters are commensurate,  the outermost regions of 
globular clusters located closer than $\sim$ 5 kpc from the LMC center seem to
have dynamically evolved faster. Having the globular clusters orbited the LMC
at different mean distances from its center along their lifetime, the distinction of their
structural properties reflect the differential tidal effects between them.

We wonder whether tidal heating still has an important role in the structural
parameters of star clusters in galaxies less massive than the LMC. 
We focus here on the Small Magellanic Cloud, which is nearly 10 times
less massive than the LMC \citep{vdmk14,stanimirovicetal2004}, because it has a statistically 
complete sample of studied star clusters to explore this issue.
\citet{gieles2007} analyzed a sample of 195 star clusters in the SMC and found
no evidence for cluster tidal dissolution in the first gigayear. They arrived to this 
conclusion by comparing the observed star cluster frequency with that predicted by 
stellar evolutionary models, assuming no tidal dissolution.

The paper is organized as follows. In Section 2 we present the data set used and 
different star cluster parameters obtained from it. Section 3 deals with the analysis
of the variation of structural parameters as a function the star cluster distance
to the SMC center. Finally, Section 4 summarizes the main results of this work.

\section{SMC star cluster properties}

We gathered information from two main sources: the recent catalog of star
clusters compiled by \citet{bicaetal2020}, from which we retrieved star cluster
ages; and Table\,2 of \citet{hz2006}, from which we used star cluster coordinates (RA 
and Dec), 90$\%$ light radii ($r_{90}$), integrated $V$ magnitudes, and
cluster eccentricities ($\epsilon$). We would like to note that different
SMC imaging surveys have been carried out since the Magellanic Clouds Photometric Survey \citep{zetal02} used by \citet{hz2006}, e.g., VMC \citep{cetal11}, 
OGLE \citep{udalskietal2015}, SMASH \citep{nideveretal2017a}, 
VISCACHA \citep{maiaetal2019}, among others. As far as we are aware, none of
these surveys have been exploited yet in order to update the parameters derived 
and analysis done by \citet{hz2006}, which justifies our choice. We computed the cluster masses using the
relationships obtained by \citet[][equation 4]{metal14} as follows:\\

log($M$ /$\msun$) = $a$ + $b$ $\times$ log(age /yr) - 0.4 $\times$ ($M_V - M_{V_{\odot}}$)\\

\noindent with $a$ = 5.87$\pm$0.07, $b$ = 0.608$\pm$0.008 for 
a representative SMC overall metallicity $Z$ = 0.004 \citep{pg13}; 
$M_{V_{\odot}}$ = 4.83.
Typical uncertainties turned out to be $\sigma$(log(M /\msun)) $\approx$ 0.2.
We note that the assumption of a single average metallicity for all star clusters 
does not affect their calculated masses, since metallicity differences imply
mass values  that are within the uncertainties 
\citep[see figures 10 and 11 in][]{metal14}. We checked that our cluster masses
are in very good agreement with those calculated by 
\citet[][see their figure 16]{hz2006}. As for the completeness of the present 
star cluster sample, we refer the reader to the work by \citet{gieles2007}, which
shows that the sample is magnitude limited.

As far as we are aware, the frequent geometry considered to analyze the spatial
distributions of SMC star clusters is the elliptical framework proposed by 
\citet{petal07d} as a simple representation of the orientation, dimension and shape of 
the SMC main body. This framework does not consider the SMC depth, which is
much more extended than the size of the galaxy projected in the sky
\citep{ripepietal2017,muravevaetal2018,graczyketal2020}. In an attempt to
represent the SMC main body more realistically, we devised a 3D geometry,
considering the SMC as an ellipsoid, as follows:

\begin{equation}
\frac{x^2}{a^2} + \frac{y^2}{b^2} + \frac{z^2}{c^2} = 1,
\end{equation}

\noindent where $x$ and $y$ directions are along the semi-minor and semi-major
axes in the \citet{petal07d}'s framework, respectively, and the $z$ axis is along the 
line-of-sight. The SMC center is adopted as the origin of this framework, i.e.,
(RA$_{SMC}$,Dec$_{SMC}$) = ($13\fdg 1875, -72\fdg 8286$)  \citep{petal07d}.
The projected ellipses in the sky have a position angle PA$_{SMC}$ = 54$\degr$ 
and a $a/b$ ratio of 1/2. 

The PAs of the star clusters in this rotated coordinate system were 
calculated using the \texttt{positionAngle} routine from \texttt{PyAstronomy}\footnote{https://github.com/sczesla/PyAstronomy} \citep[PyA,][]{pya}, 
and the observed distances in the sky to the SMC center in R.A. ($x_0$) and
Dec. ($y_0$), respectively, as follows:\\

$x_0$ = -(RA - RA$_{SMC}$)  cos(Dec) cos(PA$_{SMC}$) + (Dec - Dec$_{SMC}$) sin(PA$_{SMC}$),\\

$y_0$ =  (RA - RA$_{SMC}$)  cos(Dec) sin(PA$_{SMC}$) + (Dec - Dec$_{SMC}$) cos(PA$_{SMC}$).\\

 We assumed that the spatial star cluster distribution
is a function of their ages \citep[see figure 8 in ][and references therein]{bicaetal2020},
so that each  ellipsoid corresponds
to a fixed age. Using the age gradient of figure 8 in \citet{bicaetal2020},  we 
entered  the star clusters' ages to estimate their corresponding semi-major axis. 
We additionally used a mean SMC distance of 62.5 kpc \citep{graczyketal2020}, and 
an average $b/c$ ratio of 1:2.3 \citep[][and references therein]{ripepietal2017,muravevaetal2018,graczyketal2020} to find the projected
distance $r = (x^2 + y^2)^{1/2}$ and $z$ values for which:

\begin{equation}
 (1 + 3 \times sin^2(PA)) \times (r/b)^2 + 5.29 \times (z/b)^2 -1 = 0,
\end{equation}

\noindent  where  $b$ (kpc) = 1.67$\times$log(age /yr) -10.85 (age $\la$ 5 Gyr) with a dispersion of
0.25 kpc representing the 95$\%$ confidence interval of the fitted parameters
\citep[figure 8 in][]{bicaetal2020}. Eq. (2) comes from eq. (1), 
$x = r \times sin(PA)$, $y = r \times cos(PA)$, $a/b$ = 1/2  and $b/c$ = 1/2.3.
We note that if we do not consider the SMC depth (z=0), then $x=x_0$ and $y=y_0$. 
The $r$ and $z$ values that
comply with eq. (2) for each star cluster were obtained 
by evaluating eq. (2) 17600 times, from a grid of values of 
$r$ from 0.0 up to 11.0 kpc, in steps of 0.1 kpc, and $z$ from 0.0 up to 16.0 kpc, in 
steps of 0.1 kpc, and then  looking for the $r$ and $z$ ones which correspond 
to the smallest of the 17600 absolute values of eq. (2), which were always smaller 
than 0.01. We note that, theoretically speaking, the resulting $r$ and $z$ value should
lead eq. (2) to be equal to zero. Finally, the linear
distance of a star cluster to the SMC center is calculated as $r_{deproj}$=  $(r^2 + z^2)^{1/2}$.  We estimated the uncertainties in $r_{deproj}$ by performing the procedure described above for 1000 realizations with $b$ values randomly chosen
in the interval [$b$-$\sigma$($b$), $b$+$\sigma$($b$)]. Then, we adopted 
$\sigma$($r_{deproj}$) = 1/2.355 times the $FWHM$ of the $r_{deproj}$ distributions,
which resulted to be typically $\approx$ 1 kpc.
Figure~\ref{fig:fig1} illustrates the projected spatial distribution
of the star cluster sample where the different deprojected distances to the SMC center
are revealed. Some star clusters projected close to the SMC center are 
relatively distance objects, while others apparently placed in the outer galaxy
regions turned out to be closer to the SMC center.

The analysis of the variation
of star cluster structural parameters as a function of their deprojected distances to
the SMC center supersedes previous ones, which are based on the star cluster
positions projected on the sky. As far as we are aware, there are very few SMC
star clusters with independent distance estimates \citep[see, e.g.][]{glattetal2008a,diasetal2016}. In general, a mean SMC distance
modulus is adopted when fitting theoretical isochrones to the CMD of a star
cluster, since changes in the distance modulus by an amount equivalent to the
average SMC depth  leads to a smaller age difference than that resulting 
from the isochrones (characterized by the same metallicity) bracketing the observed 
star cluster features in the CMD. Nevertheless, there is still differences between
individual star cluster estimates. \citet{glattetal2008a} estimated distances for
NGC\,121, Lindsay\,1 and Kron\,3 of 64.9$\pm$1.2 kpc, 56.9$\pm$1.0 kpc and
60.4$\pm$1.1, respectively. However, \citet{cetal01} obtained 59.6$\pm$1.8 kpc,
53.2$\pm$0.9 kpc and 56.7$\pm$1.9 kpc, respectively.

\begin{figure}
\includegraphics[width=\columnwidth]{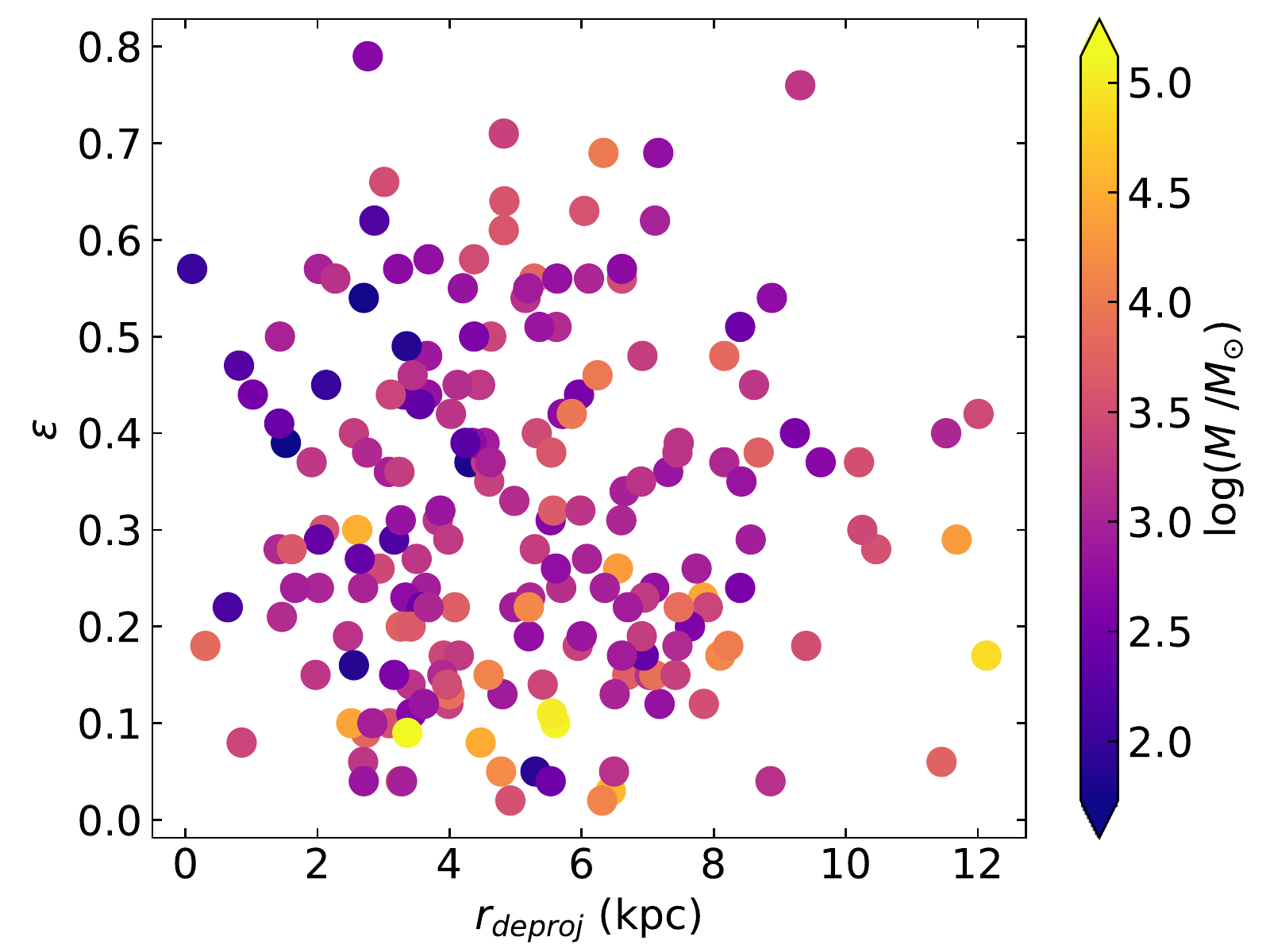}
\caption{Star cluster eccentricity versus deprojected distance from the SMC center,
color-coded according to the star cluster mass.}
\label{fig:fig2}
\end{figure}

\section{Analysis and discussion}

The different gravitational field strengths experienced by star clusters affect their structural parameters, and ultimately their internal dynamical evolutionary stages.
For example, the increase of core, half-mass, and Jacobi radii as a function of the  star
cluster distance from the Milky Way center was predicted theoretically by \citet{hm2010}
and \citet{bianchinietal2015}, among others. Star clusters in weaker tidal fields, 
like those located in the outermost regions of the Milky Way can expand naturally, 
while those immersed in stronger tidal fields (e.g. the Milky Way bulge) do not. 
We here use the calculated deprojected distances as a proxy of the  SMC  
gravitational field, to investigate whether some star cluster properties show any
trend with it.

\begin{figure}
\includegraphics[width=\columnwidth]{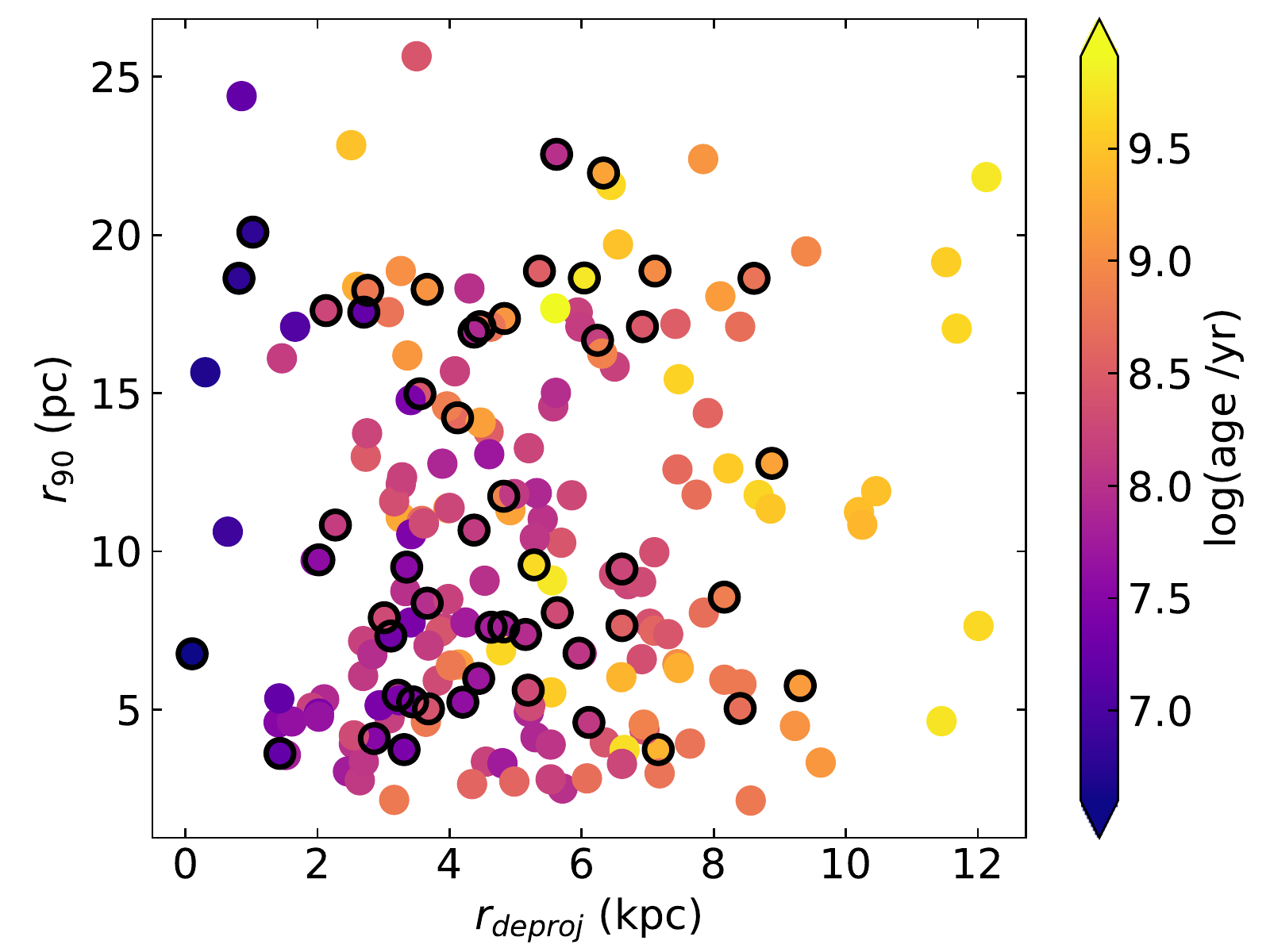}
\caption{Star cluster size ($r_{90}$)  versus deprojected distance from the SMC 
center, color-coded according to their ages. Star clusters with $\epsilon$ $>$ 0.4 are highlighted with black open circles.}
\label{fig:fig3}
\end{figure}

Figure~\ref{fig:fig2} shows the eccentricity versus deprojected distance plane for the
studied star cluster sample, from which some obvious features arise at a glance. 
The eccentricities
span a wide range of values (0.0 $<$ $\epsilon$ $<$ 0.8) for deprojected distances 
$\la$ 7-8 kpc from the SMC center. For larger deprojected distances, they span a
significantly narrower range (0.0 $<$ $\epsilon$ $\la$ 0.4).  This behavior seems
to be independent of the star cluster size, because relatively small and large
objects are located throughout the whole covered SMC body (see Fig.~\ref{fig:fig3}). 
The morphology of 
star  clusters can be shaped by different mechanisms, such as dynamical
relaxation and decay of initial velocity anisotropy, cluster rotation, external
perturbations, differential interstellar extinction, etc \citep[see][for a review]{chch2010}.
Milky Way globular clusters have a median eccentricity of $\sim$ 0.13, with those
close to the galaxy bulge having various degrees of flattening, in comparison
with those away from the Galactic center that tend to be spherical. In the LMC, the
globular cluster population shows evidence for radial variation of the cluster
eccentricity \citep{kontizasetal1989}, while in the SMC \citet{hz2006} find that the eccentricity of
star clusters correlates with their masses more strongly than with their ages. 
Figure~\ref{fig:fig2} reveals that the correlation of the eccentricity with the star cluster
mass is not apparent, because star clusters less massive than log($M$ /$\msun$)
$\sim 4.0$ are distributed at any eccentricity. However, there is a hint for more 
massive star clusters to have in general terms eccentricities smaller than 0.4. 
This would make massive SMC star clusters to belong to a distinct group of objects.

The two different eccentricity regimes mentioned above (for $r_{deproj}$ smaller or larger than $\sim$ 7-8 kpc) would also seem to be a distinguished feature. We note
here that because of the  existence of an age gradient, these two eccentricity 
regimes could  hide an eccentricity-age dependence. 
 
The trend of star cluster ages with the deprojected distances is observed in
Fig.~\ref{fig:fig3}, where some correlation arises, in the sense that the older the star
cluster the farther its location to the SMC center. However, the oldest star clusters
are not  the most distant ones to the SMC center, but somehow located at the midst 
of  the deprojected distance range, where young star clusters are also seen. Such a 
mixture of star cluster ages along the deprojected distances is caused by the 
spheroidal geometry adopted to map
more tightly the observed SMC structure and star cluster age gradient. For example,
the plane $z$ = 0.0 kpc contains old star clusters (the outermost ones in the
plane of the sky), that are located comparatively closer to the SMC center than
younger star clusters observed along the line-of-sight.

\begin{figure}
\includegraphics[width=\columnwidth]{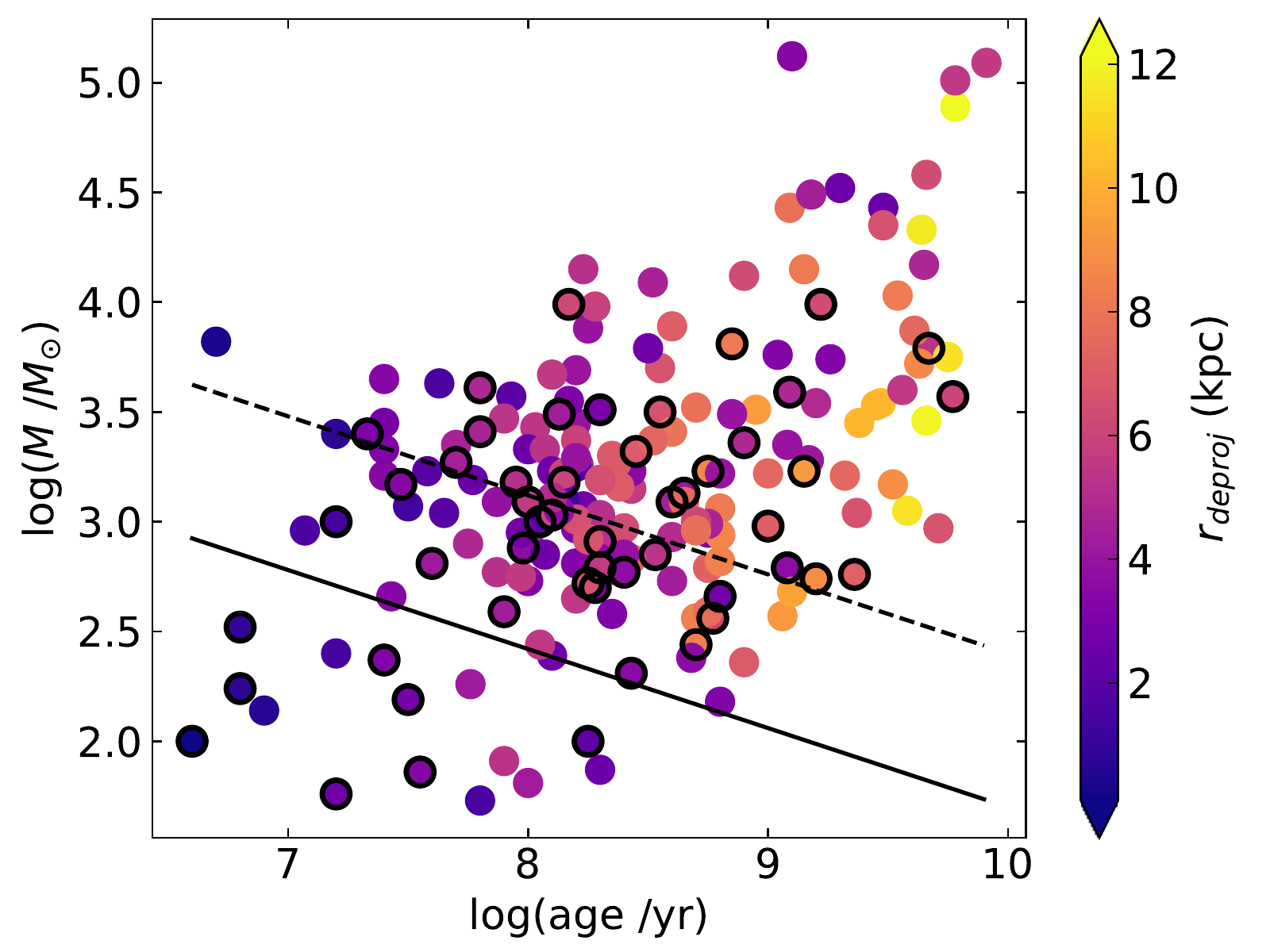}
\caption{Star cluster mass versus deprojected distance from the SMC 
center, color-coded according to their deprojected distances from the SMC center. 
Star clusters with $\epsilon$ $>$ 0.4 are highlighted with black open circles. The
straight solid line is the relationship found by \citet{joshietal2016} for Milky Way open 
clusters located within 1.8 kpc from the Sun, while the dashed and dotted ones
are parallel ones to that of \citet{joshietal2016} drawn for comparison purposes
(see text for details).}
\label{fig:fig4}
\end{figure}

Star clusters with eccentricities larger than $\sim$ 0.4 nearly span the whole age range,
as those with smaller eccentricities do (see also Fig.~\ref{fig:fig4}). This is visible 
from the inspection of Fig.~\ref{fig:fig3} for deprojected distances smaller than $\sim$ 7-8 kpc. Therefore,
an eccentricity-age dependence does not find any support. This result is not 
in opposition with the fact that star clusters with eccentricities smaller than
$\sim$ 0.4 and located at deprojected distances larger than $\sim$ 7-8 kpc are 
among the old SMC star clusters. It would seem that there is a group of massive and 
old star clusters located in the outermost SMC regions with relatively small 
eccentricities, rather than a correlation of the eccentricity with the star cluster
mass and age, 

Figure~\ref{fig:fig3} also tells us that the star cluster sizes do not show any
correlation with the deprojected distances, i.e., they would not be affected
by the SMC gravitation field, as it is the case of Milky Way and LMC globular
clusters \citep{pm2018,piattietal2019b}, which are bigger as they are located
further from the galaxy center. This finding puts a limit to the galaxy mass,
a value in between the LMC and the SMC mass, in order to  the galaxy gravitational 
field can drive the size of its star clusters. We point out that
old globular clusters in the Milky Way and the LMC are on average one
order of magnitude more massive than  massive SMC star clusters \citep{pm2018}, 
so that the comparison between them could favor a minimum galaxy mass more 
similar to that of the LMC. This also could have its own impact in the computation of
the cluster mass lost by tidal disruption along the entire lifetime of star clusters
stripped off the SMC by the LMC \citep{carpinteroetal2013}. 
In the standard cosmological scenario \citep{mooreetal1999,dl2008}, accreted 
globular clusters are formed in small dwarf galaxies. Hence, most of the cluster mass 
lost by tidal disruption should have disrupted once the star cluster is under the
effects of the Milky Way gravitational field, because low mass galaxies would not seem to affect seriously the mass budget of its massive globular clusters.
 Nevertheless, the large eccentricity values found only in SMC star
clusters located inside a volume of radius $\sim$ 7-8 kpc, implies some kind
of distortion that might be caused by the SMC tidal forces. At this point, it is
a conundrum that many star clusters distributed in a similar volume do not
have large eccentricities (see also Fig.~\ref{fig:fig2}). We also point out that
$r_{90}$, although a robust estimate of the star cluster size, does not represent
the cluster Jacobi radius, which should strictly speaking be considered for monitoring
any change in the star cluster dimension with the deprojected distance.
Typical errors in $r_{90}$ are $\sim$ 30$\%$.

The mass versus age diagram of SMC star clusters depicted in Fig.~\ref{fig:fig4}
shows that those with eccentricities larger than $\sim$ 0.4 are less massive
than  log($M$ /$\msun$) $\sim 4.0$. More massive star clusters have eccentricities
smaller than $\sim$ 0,4 and seem to be among the oldest objects. We note however
that not every old star cluster is more massive than log($M$ /$\msun$) $\sim 4.0$. Likewise, we wonder on the presence of many star clusters
less massive than log($M$ /$\msun$) $\sim 4.0$ with eccentricities smaller than
$\sim$ 0.4. Some aspects worthy of consideration to find an explanation, 
although beyond the scope of the present data sets, could be the existence of families 
of star clusters with different rotation velocities, or a differential perturbation by 
the LMC during the last close passage to the SMC \citep{pateletal2020}.

Figure~\ref{fig:fig4} also shows that the cluster mass distribution as a function of
age is quite different from that of Milky Way open clusters located in a circle of
radius 1.8 kpc from the Sun \citep[][solid line]{joshietal2016}. In the case of these
open clusters, we can assume that the mass variation as a function of their ages 
is mainly caused by evolutionary effects, if the Milky Way gravitation field does not 
affect differently them in that relatively small circle. Furthermore, we can
imagine straight lines parallel to that for \citet{joshietal2016}'s open clusters 
that correspond to star clusters under different tidal disruption regimes
 \citep{piattietal2019b}, with those for weaker tidal 
fields located upward. Figure~\ref{fig:fig4} shows a large number of SMC clusters that
would seem to follow a similar trend, shifted by $\Delta$(log($M$ /$\msun$))  
$\sim$ 0.7 (dashed line)  toward larger masses. This nearly constant 
log mass  difference could reflect the much stronger tidal field of the Milky Way
at the solar circle in comparison with that of the SMC, assumed that the SMC star clusters are affected by the same SMC tidal field strength. We note that such
a trend is followed by star clusters with some hundred Myr, for which  mass loss
is mainly driven by stellar evolution, and also for some older star clusters,
where two-body relaxation can have a more important role. Star clusters older
than $\sim$ 1 Gyr practically did not survive along the dashed line. However,
if more massive star clusters had experienced mass loss by tidal disruption as
those along the dashed line, some of them would have been seen populating
the region around the dashed line (log(age /yr) $\ga$ 9.3). The fact that old
clusters appear  above the dashed line could be interpreted as they 
are affected by weaker gravitational field strengths. 
We note that most of them have eccentricities $\la$ 0.4, and are located at deprojected distances $\ga$ 7-8 kpc.
The observed mass range at any age is  $\Delta$(log($M$ /$\msun$))  $\sim$ 2.0.

\section{Concluding remarks}

We made use of available data sets of structural properties for a statistically 
significant sample of SMC star clusters with the aim of studying at what extend
the SMC gravitational field are responsible of the star cluster shapes and sizes.
Recently, it was shown the observed dependence of the core, half-mass, and
Jacobi radii, alongside relaxation time, cluster mass loss by tidal disruption, 
among others, with the position in the galaxy of old globular Milky Way and LMC 
clusters. Although the SMC does not harbor star clusters as old as the ancient
globular clusters, the spatial coverage of star clusters spanning the whole age
range allows us to probe for tidal effects. \citet{hz2006} performed an analysis
of some structural properties of SMC star clusters. As far as we are aware,
this is the first time that star cluster properties are analyzed in the context of the
3D geometry of the SMC.

We adopted an ellipsoid as a representation of the SMC with the three axes
having different extensions. They have been adjusted combining the known star
cluster age gradient and the recently SMC depth estimated from Classical
Cepheids, RR Lyrae stars, and late-type eclipsing binaries. In this framework,
each age is assigned to a unique ellipsoid. Therefore, by using the age of the
star clusters and their projected positions in the sky with respect to the SMC center,
we estimated their deprojected distances, which we used as a proxy of the
SMC gravitational field. The use of deprojected distances  solved
the spurious effect of considering a star cluster to be located close to the SMC center,
from its projected position in the sky.

We sought any trend between the star cluster size (represented by the 90$\%$ 
light radius), the eccentricity, the mass and age with the deprojected distance. 
We did find that the size of the star clusters would not seem to be sensitive to
changes in their positions in the galaxy, because star clusters spanning the entire
observed range are found everywhere. We point out, however, that Jacobi radii
would be appropriate for a more definitive answer. The star cluster eccentricities
reveal that those more elongated objects ($\epsilon$ $\ga$ 0.4) are preferencially  
located at deprojected distances $\la$ 7-8 kpc. This finding could be a hint for
differential tidal effects between star clusters located closer and farther from the
SMC center. However, we found a numerous population of stars clusters 
distributed inside the same volume that look like less elongated ($\epsilon$ $\la$ 0.4).

Star clusters with estimated masses larger than log($M$ /$M_{\odot}$) $\sim$ 4.0
have relatively small eccentricities ($\epsilon$ $\la$ 0.4), are older than log(age /yr) 
$\sim$ 9.0, considering the uncertainties in their estimated masses, and are mostly located in the outermost regions of the SMC.  We would like to remind that
we initially assumed a dependence in deprojected distance and cluster mass on age. These features could favor an 
scenario of differential tidal effects. Likewise, there is a large number of star
clusters located at deprojected distances $\la$ 7-8 kpc that mimic the linear 
cluster mass versus age relationship of Milky Way open clusters placed within a circle
of radius 1.8 kpc from the Sun, with a zero point offset of 0.7 toward more massive
star clusters. We interpret this shift as originating from different gravitational field
strengths.

\begin{acknowledgements}
 I thank the referee for the thorough reading of the manuscript and
timely suggestions to improve it. 
\end{acknowledgements}


\begin{thebibliography}{47}
\expandafter\ifx\csname natexlab\endcsname\relax\def\natexlab#1{#1}\fi

\bibitem[{{Baumgardt} \& {Makino}(2003)}]{bm2003}
{Baumgardt}, H. \& {Makino}, J. 2003, \mnras, 340, 227

\bibitem[{{Bianchini} {et~al.}(2015){Bianchini}, {Renaud}, {Gieles}, \&
  {Varri}}]{bianchinietal2015}
{Bianchini}, P., {Renaud}, F., {Gieles}, M., \& {Varri}, A.~L. 2015, \mnras,
  447, L40

\bibitem[{{Bica} {et~al.}(2020){Bica}, {Westera}, {Kerber}, {Dias}, {Maia},
  {Santos}, {Barbuy}, \& {Oliveira}}]{bicaetal2020}
{Bica}, E., {Westera}, P., {Kerber}, L. d.~O., {et~al.} 2020, \aj, 159, 82

\bibitem[{{Carpintero} {et~al.}(2013){Carpintero}, {G{\'o}mez}, \&
  {Piatti}}]{carpinteroetal2013}
{Carpintero}, D.~D., {G{\'o}mez}, F.~A., \& {Piatti}, A.~E. 2013, \mnras, 435
  [\eprint[arXiv]{1307.6231}]

\bibitem[{{Chen} \& {Chen}(2010)}]{chch2010}
{Chen}, C.~W. \& {Chen}, W.~P. 2010, \apj, 721, 1790

\bibitem[{{Cioni} {et~al.}(2011){Cioni}, {Clementini}, {Girardi}, {Guandalini},
  {Gullieuszik}, {Miszalski}, {Moretti}, {Ripepi}, {Rubele}, {Bagheri},
  {Bekki}, {Cross}, {de Blok}, {de Grijs}, {Emerson}, {Evans}, {Gibson},
  {Gonzales-Solares}, {Groenewegen}, {Irwin}, {Ivanov}, {Lewis}, {Marconi},
  {Marquette}, {Mastropietro}, {Moore}, {Napiwotzki}, {Naylor}, {Oliveira},
  {Read}, {Sutorius}, {van Loon}, {Wilkinson}, \& {Wood}}]{cetal11}
{Cioni}, M.-R.~L., {Clementini}, G., {Girardi}, L., {et~al.} 2011, \aap, 527,
  A116

\bibitem[{{Crowl} {et~al.}(2001){Crowl}, {Sarajedini}, {Piatti}, {Geisler},
  {Bica}, {Clari{\'a}}, \& {Santos}}]{cetal01}
{Crowl}, H.~H., {Sarajedini}, A., {Piatti}, A.~E., {et~al.} 2001, \aj, 122, 220

\bibitem[{{Czesla} {et~al.}(2019){Czesla}, {Schr{\"o}ter}, {Schneider},
  {Huber}, {Pfeifer}, {Andreasen}, \& {Zechmeister}}]{pya}
{Czesla}, S., {Schr{\"o}ter}, S., {Schneider}, C.~P., {et~al.} 2019, {PyA:
  Python astronomy-related packages}

\bibitem[{{Deason} {et~al.}(2020){Deason}, {Erkal}, {Belokurov}, {Fattahi},
  {G{\'o}mez}, {Grand }, {Pakmor}, {Xue}, {Liu}, {Yang}, {Zhang}, \&
  {Zhao}}]{deasonetal2020}
{Deason}, A.~J., {Erkal}, D., {Belokurov}, V., {et~al.} 2020, arXiv e-prints,
  arXiv:2010.13801

\bibitem[{{Dias} {et~al.}(2016){Dias}, {Kerber}, {Barbuy}, {Bica}, \&
  {Ortolani}}]{diasetal2016}
{Dias}, B., {Kerber}, L., {Barbuy}, B., {Bica}, E., \& {Ortolani}, S. 2016,
  \aap, 591, A11

\bibitem[{{D'Onghia} \& {Lake}(2008)}]{dl2008}
{D'Onghia}, E. \& {Lake}, G. 2008, \apjl, 686, L61

\bibitem[{{Gieles} \& {Baumgardt}(2008)}]{gielesetal2008}
{Gieles}, M. \& {Baumgardt}, H. 2008, \mnras, 389, L28

\bibitem[{{Gieles} {et~al.}(2011){Gieles}, {Heggie}, \&
  {Zhao}}]{gielesetal2011}
{Gieles}, M., {Heggie}, D.~C., \& {Zhao}, H. 2011, \mnras, 413, 2509

\bibitem[{{Gieles} {et~al.}(2007){Gieles}, {Lamers}, \& {Portegies
  Zwart}}]{gieles2007}
{Gieles}, M., {Lamers}, H.~J.~G.~L.~M., \& {Portegies Zwart}, S.~F. 2007, \apj,
  668, 268

\bibitem[{{Gieles} {et~al.}(2006){Gieles}, {Portegies Zwart}, {Baumgardt},
  {Athanassoula}, {Lamers}, {Sipior}, \& {Leenaarts}}]{gielesetal2006}
{Gieles}, M., {Portegies Zwart}, S.~F., {Baumgardt}, H., {et~al.} 2006, \mnras,
  371, 793

\bibitem[{{Gieles} \& {Renaud}(2016)}]{gr2016}
{Gieles}, M. \& {Renaud}, F. 2016, \mnras, 463, L103

\bibitem[{{Glatt} {et~al.}(2008){Glatt}, {Grebel}, {Sabbi}, {Gallagher},
  {Nota}, {Sirianni}, {Clementini}, {Tosi}, {Harbeck}, {Koch}, {Kayser}, \& {Da
  Costa}}]{glattetal2008a}
{Glatt}, K., {Grebel}, E.~K., {Sabbi}, E., {et~al.} 2008, \aj, 136, 1703

\bibitem[{{Gnedin} {et~al.}(1999){Gnedin}, {Lee}, \&
  {Ostriker}}]{gnedinetal1999}
{Gnedin}, O.~Y., {Lee}, H.~M., \& {Ostriker}, J.~P. 1999, \apj, 522, 935

\bibitem[{{Graczyk} {et~al.}(2020){Graczyk}, {Pietrzynski}, {Thompson},
  {Gieren}, {Zgirski}, {Villanova}, {Gorski}, {Wielgorski}, {Karczmarek},
  {Narloch}, {Pilecki}, {Taormina}, {Smolec}, {Suchomska}, {Gallenne},
  {Nardetto}, {Storm}, {Kudritzki}, {Kaluszynski}, \& {Pych}}]{graczyketal2020}
{Graczyk}, D., {Pietrzynski}, G., {Thompson}, I.~B., {et~al.} 2020, arXiv
  e-prints, arXiv:2010.08754

\bibitem[{{Heggie} \& {Hut}(2003)}]{hh03}
{Heggie}, D. \& {Hut}, P. 2003, {The Gravitational Million-Body Problem: A
  Multidisciplinary Approach to Star Cluster Dynamics}

\bibitem[{{Hill} \& {Zaritsky}(2006)}]{hz2006}
{Hill}, A. \& {Zaritsky}, D. 2006, \aj, 131, 414

\bibitem[{{Hurley} \& {Mackey}(2010)}]{hm2010}
{Hurley}, J.~R. \& {Mackey}, A.~D. 2010, \mnras, 408, 2353

\bibitem[{{Joshi} {et~al.}(2016){Joshi}, {Dambis}, {Pandey}, \&
  {Joshi}}]{joshietal2016}
{Joshi}, Y.~C., {Dambis}, A.~K., {Pandey}, A.~K., \& {Joshi}, S. 2016, \aap,
  593, A116

\bibitem[{{Kontizas} {et~al.}(1989){Kontizas}, {Kontizas}, {Sedmak}, \&
  {Smareglia}}]{kontizasetal1989}
{Kontizas}, E., {Kontizas}, M., {Sedmak}, G., \& {Smareglia}, R. 1989, \aj, 98,
  590

\bibitem[{{Kruijssen} \& {Mieske}(2009)}]{km2009}
{Kruijssen}, J.~M.~D. \& {Mieske}, S. 2009, \aap, 500, 785

\bibitem[{{Kruijssen} {et~al.}(2011){Kruijssen}, {Pelupessy}, {Lamers},
  {Portegies Zwart}, \& {Icke}}]{kruijssenetal2011}
{Kruijssen}, J.~M.~D., {Pelupessy}, F.~I., {Lamers}, H. J.~G.~L.~M., {Portegies
  Zwart}, S.~F., \& {Icke}, V. 2011, \mnras, 414, 1339

\bibitem[{{Lamers} \& {Gieles}(2006)}]{lg2006}
{Lamers}, H.~J.~G.~L.~M. \& {Gieles}, M. 2006, \aap, 455, L17

\bibitem[{{Lamers} {et~al.}(2005){Lamers}, {Gieles}, {Bastian}, {Baumgardt},
  {Kharchenko}, \& {Portegies Zwart}}]{lamersetal2005a}
{Lamers}, H.~J.~G.~L.~M., {Gieles}, M., {Bastian}, N., {et~al.} 2005, \aap,
  441, 117

\bibitem[{{Maia} {et~al.}(2019){Maia}, {Dias}, {Santos}, {Kerber}, {Bica},
  {Piatti}, {Barbuy}, {Quint}, {Fraga}, {Sanmartim}, {Angelo},
  {Hernandez-Jimenez}, {Katime Santrich}, {Oliveira}, {P{\'e}rez-Villegas},
  {Souza}, {Vieira}, \& {Westera}}]{maiaetal2019}
{Maia}, F.~F.~S., {Dias}, B., {Santos}, J.~F.~C., {et~al.} 2019, \mnras, 484,
  5702

\bibitem[{{Maia} {et~al.}(2014){Maia}, {Piatti}, \& {Santos}}]{metal14}
{Maia}, F.~F.~S., {Piatti}, A.~E., \& {Santos}, J.~F.~C. 2014, \mnras, 437,
  2005

\bibitem[{{Moore} {et~al.}(1999){Moore}, {Ghigna}, {Governato}, {Lake},
  {Quinn}, {Stadel}, \& {Tozzi}}]{mooreetal1999}
{Moore}, B., {Ghigna}, S., {Governato}, F., {et~al.} 1999, \apjl, 524, L19

\bibitem[{{Muraveva} {et~al.}(2018){Muraveva}, {Subramanian}, {Clementini},
  {Cioni}, {Palmer}, {van Loon}, {Moretti}, {de Grijs}, {Molinaro}, {Ripepi},
  {Marconi}, {Emerson}, \& {Ivanov}}]{muravevaetal2018}
{Muraveva}, T., {Subramanian}, S., {Clementini}, G., {et~al.} 2018, \mnras,
  473, 3131

\bibitem[{{Nidever} {et~al.}(2017){Nidever}, {Olsen}, {Walker}, {Vivas},
  {Blum}, {Kaleida}, {Choi}, {Conn}, {Gruendl}, {Bell}, {Besla}, {Mu{\~n}oz},
  {Gallart}, {Martin}, {Olszewski}, {Saha}, {Monachesi}, {Monelli}, {de Boer},
  {Johnson}, {Zaritsky}, {Stringfellow}, {van der Marel}, {Cioni}, {Jin},
  {Majewski}, {Martinez-Delgado}, {Monteagudo}, {No{\"e}l}, {Bernard},
  {Kunder}, {Chu}, {Bell}, {Santana}, {Frechem}, {Medina}, {Parkash},
  {Navarrete}, \& {Hayes}}]{nideveretal2017a}
{Nidever}, D.~L., {Olsen}, K., {Walker}, A.~R., {et~al.} 2017, \aj, 154, 199

\bibitem[{{Patel} {et~al.}(2020){Patel}, {Kallivayalil}, {Garavito-Camargo},
  {Besla}, {Weisz}, {van der Marel}, {Boylan-Kolchin}, {Pawlowski}, \&
  {G{\'o}mez}}]{pateletal2020}
{Patel}, E., {Kallivayalil}, N., {Garavito-Camargo}, N., {et~al.} 2020, \apj,
  893, 121

\bibitem[{{Piatti}(2019)}]{piatti2019}
{Piatti}, A.~E. 2019, \apj, 882, 98

\bibitem[{{Piatti} \& {Geisler}(2013)}]{pg13}
{Piatti}, A.~E. \& {Geisler}, D. 2013, \aj, 145, 17

\bibitem[{{Piatti} \& {Mackey}(2018)}]{pm2018}
{Piatti}, A.~E. \& {Mackey}, A.~D. 2018, \mnras, 478, 2164

\bibitem[{{Piatti} {et~al.}(2007){Piatti}, {Sarajedini}, {Geisler}, {Clark}, \&
  {Seguel}}]{petal07d}
{Piatti}, A.~E., {Sarajedini}, A., {Geisler}, D., {Clark}, D., \& {Seguel}, J.
  2007, \mnras, 377, 300

\bibitem[{{Piatti} {et~al.}(2019){Piatti}, {Webb}, \&
  {Carlberg}}]{piattietal2019b}
{Piatti}, A.~E., {Webb}, J.~J., \& {Carlberg}, R.~G. 2019, \mnras, 489, 4367

\bibitem[{{Ripepi} {et~al.}(2017){Ripepi}, {Cioni}, {Moretti}, {Marconi},
  {Bekki}, {Clementini}, {de Grijs}, {Emerson}, {Groenewegen}, {Ivanov},
  {Molinaro}, {Muraveva}, {Oliveira}, {Piatti}, {Subramanian}, \& {van
  Loon}}]{ripepietal2017}
{Ripepi}, V., {Cioni}, M.-R.~L., {Moretti}, M.~I., {et~al.} 2017, \mnras, 472,
  808

\bibitem[{{Shukirgaliyev} {et~al.}(2018){Shukirgaliyev}, {Parmentier}, {Just},
  \& {Berczik}}]{shukirgaliyevetal2018}
{Shukirgaliyev}, B., {Parmentier}, G., {Just}, A., \& {Berczik}, P. 2018, \apj,
  863, 171

\bibitem[{{Stanimirovi{\'c}} {et~al.}(2004){Stanimirovi{\'c}},
  {Staveley-Smith}, \& {Jones}}]{stanimirovicetal2004}
{Stanimirovi{\'c}}, S., {Staveley-Smith}, L., \& {Jones}, P.~A. 2004, \apj,
  604, 176

\bibitem[{{Udalski} {et~al.}(2015){Udalski}, {Szyma{\'n}ski}, \&
  {Szyma{\'n}ski}}]{udalskietal2015}
{Udalski}, A., {Szyma{\'n}ski}, M.~K., \& {Szyma{\'n}ski}, G. 2015, \actaa, 65,
  1

\bibitem[{{van der Marel} \& {Kallivayalil}(2014)}]{vdmk14}
{van der Marel}, R.~P. \& {Kallivayalil}, N. 2014, \apj, 781, 121

\bibitem[{{Webb} {et~al.}(2013){Webb}, {Harris}, {Sills}, \&
  {Hurley}}]{webbetal2013}
{Webb}, J.~J., {Harris}, W.~E., {Sills}, A., \& {Hurley}, J.~R. 2013, \apj,
  764, 124

\bibitem[{{Webb} {et~al.}(2014){Webb}, {Sills}, {Harris}, \&
  {Hurley}}]{webbetal2014}
{Webb}, J.~J., {Sills}, A., {Harris}, W.~E., \& {Hurley}, J.~R. 2014, \mnras,
  445, 1048

\bibitem[{{Zaritsky} {et~al.}(2002){Zaritsky}, {Harris}, {Thompson}, {Grebel},
  \& {Massey}}]{zetal02}
{Zaritsky}, D., {Harris}, J., {Thompson}, I.~B., {Grebel}, E.~K., \& {Massey},
  P. 2002, \aj, 123, 855

\end{thebibliography}


\end{document}